\documentclass[11pt,a4paper]{article}

\usepackage{a4wide}
\usepackage[pdfborderstyle={/S/S/W 1}]{hyperref}
\usepackage[compress]{cite}
\usepackage{tikz}
\usetikzlibrary{arrows,shapes}
\usepackage{multirow}

\newcommand{\figref}[1]{Figure~\ref{fig:#1}}
\newcommand{\tabref}[1]{Table~\ref{tab:#1}}

\newcommand{\biomodels}[1]{\href{http://identifiers.org/biomodels.db/#1}{#1}}
\newcommand{\chebi}[2]{\href{http://identifiers.org/chebi/CHEBI:#2}{#1}}
\newcommand{\doi}[1]{\href{http://dx.doi.org/#1}{doi:#1}}
\newcommand{\isbn}[1]{\href{http://identifiers.org/isbn/#1}{isbn:#1}}

\title{Striking a balance with Recon 2.1}
\author{Kieran Smallbone \\ [24pt]
Manchester Centre for Integrative Systems Biology\\
131 Princess Street, Manchester M1 7DN, UK\\
\href{mailto:kieran.smallbone@manchester.ac.uk}{\tt kieran.smallbone@manchester.ac.uk}
}

\date{}

\begin{document}
\maketitle

\begin{abstract}
\noindent Recon 2 is a highly curated reconstruction of the human metabolic network. Whilst the network is state of the art, it has shortcomings, including the presence of unbalanced reactions involving generic metabolites. By replacing these generic molecules with each of their specific instances, we can ensure full elemental balancing, in turn allowing constraint-based analyses to be performed. The resultant model, called Recon 2.1, is an order of magnitude larger than the original.
\end{abstract}

\section*{Introduction}

A comprehensive, curated, consensus reconstruction of the human metabolic network, known as ``Recon 2'', was released in 2013~\cite{recon2}. Its development followed a jamboree approach~\cite{yeast1, jamborees}, exploiting both genomic and literature data to expand upon existing models~\cite{recon1, ehmn, hepatonet}.

The reconstruction is primarily described and made available in SBML (Systems Biology Markup Language, \href{http://sbml.org/}{http://sbml.org/}), an established community XML format for the mark-up of biochemical models that is understood by a large number of software applications~ \cite{sbml}. Cellular compartments, metabolites, genes and enzymes are annotated using ontological terms. These annotations provide the facility to assign definitive terms to individual components, allowing  software to identify such components unambiguously and thus link model components to existing data resources~\cite{kell08}. MIRIAM--compliant annotations have been used to identify components unambiguously by associating them with one or more terms from publicly available databases~\cite{miriam1, miriam2}. Thus this network is entirely traceable and is presented as a computational framework.

In contrast to other resources such as KEGG~\cite{kegg}, Recon 2 acts not only as a knowledge-base, but also as a predictive model, amenable to constraint-based approaches such as flux balance analysis~\cite{orth10}. Whilst it represents the state of the art of public human metabolic network reconstructions, it does have known shortcomings~\cite{swainston13}. It is anticipated that the reconstruction will undergo a cycle of iterative improvement and release, following the strategy adopted in baker's yeast~\cite{yeast1, yeast4, yeast5, yeast6, yeast7}.

\section*{Carbon balancing}

\begin{figure}[!p]
	\centering
    \resizebox{\textwidth}{!}{
      \begin{tikzpicture}
        [thick,
        r/.style={circle, draw, minimum size=1cm},
        c16/.style={r, font={C$_{16}$}},
        c14/.style={r, font={C$_{14}$}},
        R/.style={r, font={R}},
        coa/.style={star, star points=8, draw, font={C$_{21}$}},
        dhap/.style={regular polygon,regular polygon sides=4, draw, font={C$_3$}},
        ]
        
        \node[R] (AA) at (0,2) {};
        \node[coa] (BB) at (2,2) {};
        \draw [-] (AA) -- (BB);
        \node at (3.5,2) {+};
        \node[dhap] at (5,2) {};
        \draw [triangle 90-triangle 90] (6.5,2) -- (8,2);
        \node[R] (CC) at (9.5,2) {};
        \node[dhap] (DD) at (11.5,2) {};
        \draw [-] (CC) -- (DD);
        \node at (13,2) {+};
        \node[coa] at (14.5,2) {};
      
        \node[c14] (A) at (3,0) {};
        \node[coa] (B) at (5,0) {};
        \draw [-] (A) -- (B);
        \draw [-triangle 90] (6.5,0) -- (8,0);
        \node at (9.5,0) {$\left(\frac{14+21}{16+21}\right)$};
        \node[R] (C) at (11,0) {};
        \node[coa] (D) at (13,0) {};
        \draw [-] (C) -- (D);
                
        \node[R] (A) at (3,-2) {};
        \node[coa] (B) at (5,-2) {};
        \draw [-] (A) -- (B);
        \draw [-triangle 90] (6.5,-2) -- (8,-2);
        \node[c16] (C) at (9.5,-2) {};
        \node[coa] (D) at (11.5,-2) {};
        \draw [-] (C) -- (D);
  
      \end{tikzpicture}
    }
	\caption{\label{fig:1}Examples of reactions found in Recon 2 involving generic metaboites. C$_3$: \chebi{dihydroxyacetone phosphate}{57642}, C$_{14}$: \chebi{myristate}{30807}, C$_{16}$: \chebi{palmitate}{7896}, C$_{21}$: \chebi{coenzyme A}{57287}, R: generic fatty acid.}
\end{figure}
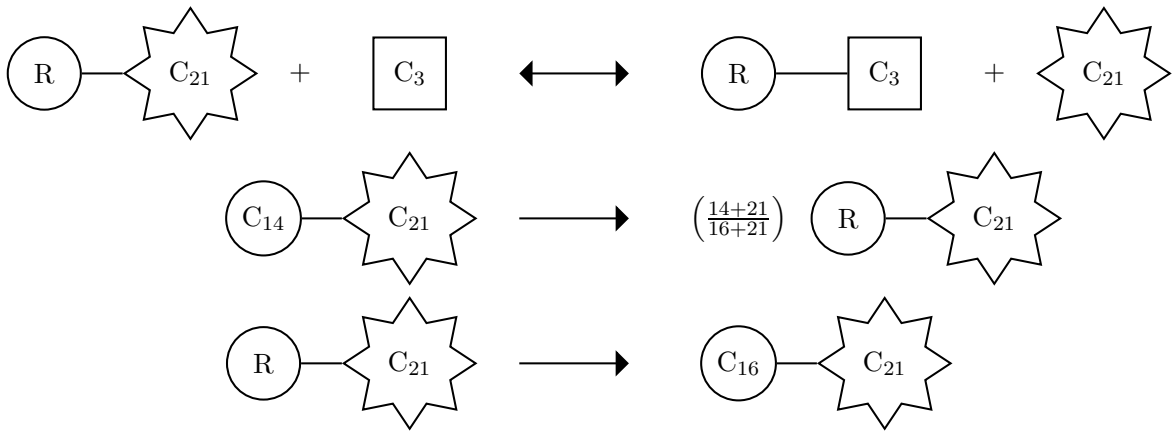

Amongst the reconstruction's problems are the presence of reactions involving generic metabolites. This can lead to elemental imbalances and violation of mass-balance. Consider \figref{1}, which presents three exemplar reactions from the model. The first reaction sees a generic fatty acid R-group transfer from a CoA to a DHAP. Such reactions provide a template which is assumed to be followed by all fatty acid side groups. However, to function as a model, there need to be connections between these generic molecules and their specific counterparts. Thus the other reactions see a specific to generic reaction, with myristoyl--CoA converted to R--CoA, and a generic-to-specific reaction, with R--CoA converted to palmitoyl--CoA.

It is clear that, taken together, the second and third reactions are unbalanced, providing a mechanism for the model to generate two carbons ``for free''. In turn, this will mean that any analyses examining carbon-limited growth, for example, will give unreliable results. Indeed Recon 2 can grow in the absence of any carbon source.

One method for overcoming this issue is to assign an (albeit arbitrary) number of carbons to each R-group. Then, through modification of the stoichiometry of the generic metabolite, each reaction may be balanced. Assigning R to contain 16 carbons (as proposed in \cite{recon1}), and changing the product stoichiometry as denoted in \figref{1}, all three equations become carbon balanced.

This methodology is applied to all reactions in the model. Various other small changes are also made, such as ensuring that all non-generic molecules are assigned a formula, and the definition of a minimal medium (see below). The resultant network, which we shall denote ``Recon 2.1'', is fully carbon balanced (see \tabref{1}) and may be used as a model in constraint-based analyses.

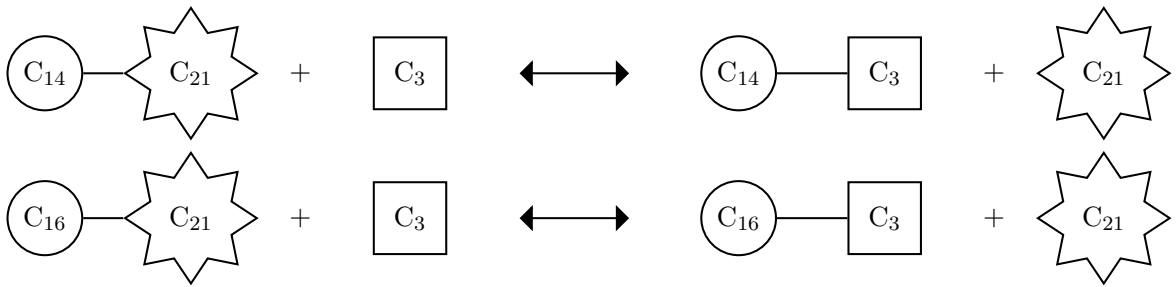
\begin{figure}[!p]
	\centering
    \resizebox{\textwidth}{!}{
      \begin{tikzpicture}
        [thick,
        r/.style={circle, draw, minimum size=1cm},
        c14/.style={r, font={C$_{14}$}},
        c16/.style={r, font={C$_{16}$}},
        R/.style={r, font={R}},
        coa/.style={star, star points=8, draw, font={C$_{21}$}},
        dhap/.style={regular polygon,regular polygon sides=4, draw, font={C$_3$}},
        ]
      
        \node[c14] (A) at (0,2) {};
        \node[coa] (B) at (2,2) {};
        \draw [-] (A) -- (B);
        \node at (3.5,2) {+};
        \node[dhap] at (5,2) {};
        \draw [triangle 90-triangle 90] (6.5,2) -- (8,2);
        \node[c14] (C) at (9.5,2) {};
        \node[dhap] (D) at (11.5,2) {};
        \draw [-] (C) -- (D);
        \node at (13,2) {+};
        \node[coa] at (14.5,2) {};
        
        \node[c16] (AA) at (0,0) {};
        \node[coa] (BB) at (2,0) {};
        \draw [-] (AA) -- (BB);
        \node at (3.5,0) {+};
        \node[dhap] at (5,0) {};
        \draw [triangle 90-triangle 90] (6.5,0) -- (8,0);
        \node[c16] (CC) at (9.5,0) {};
        \node[dhap] (DD) at (11.5,0) {};
        \draw [-] (CC) -- (DD);
        \node at (13,0) {+};
        \node[coa] at (14.5,0) {};
          
      \end{tikzpicture}
    }
	\caption{\label{fig:2}Specific versions of the generic reactions in \figref{1}.}
\end{figure}

\subsection*{Minimal medium}

A minimal set of nutrients is defined on which the cell is allowed to grow: \chebi{glucose}{4167}, \chebi{ammonium}{28938}, \chebi{oxygen}{15379}, \chebi{phosphate}{43474} and \chebi{sulphate}{16189}; the essential amino acids \chebi{histidine}{15971}, \chebi{isoleucine}{17191}, \chebi{leucine}{15603}, \chebi{lysine}{32551}, \chebi{methionine}{16643}, \chebi{phenylalanine}{17295}, \chebi{threonine}{16857}, \chebi{tryptophan}{16828}, and \chebi{valine}{16414}; the essential fatty acids \chebi{$\alpha$-linoleate}{32387} and \chebi{linoleate}{30245}; dietary minerals of \chebi{calcium}{29108}, \chebi{chloride}{17996}, \chebi{iron}{29033}, \chebi{potassium}{29103}, and \chebi{sodium}{29101}; vitamins \chebi{${\rm B_2}$}{17015} and \chebi{${\rm B_5}$}{7916}; and \chebi{protons}{24636} and \chebi{water}{15377}.

Typically, analyses (for example, those presented in \cite{shlomi11}) would proceed based on there being a single limiting nutrient, which leads to a constant growth yield. Here, however, the model may use glucose or fatty acids as carbohydrate sources, and thus may grow without any glucose present. Instead of limiting glucose uptake, we limit the total carbon uptake of the cell. From a practical perspective, this is achieved by first splitting each exchange reaction into its import and export parts. A carbon meta-reactant in then added to import reactions with stoichiometry equal to the number of carbons of that metabolite. For example:

$$\rm glucose_{\, boundary} + 6~carbons \longrightarrow glucose_{\, extracellular}$$

The total carbon uptake limit is then set to six flux units, equivalent to a typical glucose uptake limit of one flux unit.

\section*{Elemental balancing}

The method outlined above creates a model that is carbon balanced. However, the same is not true of other elements. To ensure that the model is elementally balanced, all generic molecules must be replaced by their specific counterparts. For the example in \figref{1}, this means creating a new reaction for each possible R-group, and removing any specific-to-generic interconversions (see \figref{2}).

A choice has to be made as to which fatty acids to include in this expansion as it may lead to a combinatorial explosion; given $n$ possible R-groups, there are $n^3$ possible triglycerides. To strike a balance between complexity and completeness, we include the eight most commonly found fatty acids in triglycerides~\cite{christie86}: \chebi{myristate}{30807} (14:0), \chebi{palmitate}{7896} (16:0), \chebi{palmitoleate}{32372} (16:1), \chebi{stearate}{25629} (18:0), \chebi{oleate}{30823} (18:1), \chebi{linoleate}{30245} (18:2), \chebi{$\alpha$-linoleate}{32387} (18:3) and \chebi{$\gamma$-linoleate}{32391} (18:3).

The methodology was applied to Recon 2.1 and the resultant model, which we shall term ``Recon 2.1x'', was significantly larger and elementally balanced (see \tabref{1}) . Two reactions were not expanded: cardiolipin synthase, as cardiolipin contains four R-groups (and thus $n^4$ possible configurations), and the biomass reaction. Instead, for generic species in these reactions only, ``is a'' reactions~\cite{yeast5} of the form of the second example in \figref{1} are added. Such reactions are not included in the statistics for \tabref{1}.

\begin{table}[!h]
	\centering
	\begin{tabular}{l l | c c c}
        &             			& Recon 2		& Recon 2.1 	& Recon 2.1x	\\
        \hline
        & reactions		& 7440		& 8089		& 71159		\\
        & variables		& 5063		& 5056		& 13940		\\        
        \hline
        \multirow{3}{*}{carbon-}
        & balanced 		& 5553		& 8089		& 71159		\\
        & unbalanced 		& 562		& 0			& 0			\\
        & unknown 		& 1325		& 0			& 0			\\
        \hline
        \multirow{3}{*}{element-}
        & balanced 		& 5504		& 7461		& 71157		\\
        & unbalanced 		& 611		& 0			& 0			\\
        & unknown 		& 1325		& 628		& 2			\\       
        \hline        
	\end{tabular}
	\caption{\label{tab:1}Statistics of the original model (Recon 2), the carbon balanced version (Recon 2.1) and the expanded, element balanced version (Recon 2.1x).}
\end{table}

\section*{Summary}

In this paper, we have modified the human metabolic reconstruction through defining a minimal medium for growth, and ensuring all reactions are at least carbon balanced. This will allow the use of the reconstruction as a model, for example in constraint-based analyses. We have also produced an elementally balanced model, through removal of all generic metabolites. This process increases the model size from around $10^4$ to $10^5$ reactions and serves as a guide to the size of model that software developers should strive to support.

We also see a three-fold increase in the number of metabolites in the model, primarily through a finer description of the lipids therein. Nonetheless, the number of unique metabolites still falls well short of the forty thousand defined in HMDB~\cite{hmdb}. We acknowledge that this work forms one further step in the iterative network reconstruction process, towards the ultimate goal of a complete description of human metabolism~\cite{reed03,bucher}.

\paragraph{Supplementary material}

The models described above are available in SBML format~\cite{sbml} from the BioModels database~\cite{biomodels}. Their accession numbers are:
\begin{itemize}
	\item Recon 2.1: \biomodels{MODEL1311110000}
	\item Recon 2.1x: \biomodels{MODEL1311110001}
\end{itemize}

\paragraph{Acknowledgements}

I am grateful for the financial support of the EU FP7 (KBBE) grant 289434 ``\href{http://www.biopredyn.eu}{BioPreDyn}: New Bioinformatics Methods and Tools for Data-Driven Predictive Dynamic Modelling in Biotechnological Applications''. Thanks to Brandon Barker, James Eddy and Emanuel Gon\c{c}alves for pointing out improvements to Recon 2, and to Neema Jamshidi for clarifying the meaning of the fatty acid description in Recon 1.

\end{document}